# Vibrational characterization of dinaphthylpolyynes:

# a model system for the study of end-capped *sp* carbon chains

*Eugenio Cinquanta[2,3], Luca Ravagnan[*,1,2], Ivano Eligio Castelli[+,1,4], Franco Cataldo [5,6]*

*Nicola Manini[1,4], Giovanni Onida[1,4], Paolo Milani[1,2]*

[1]Dipartimento di Fisica, Università degli Studi di Milano, Via Celoria 16, I-20133 Milano, Italy

[2]CIMAINA, Università degli Studi di Milano, Via Celoria 16, I-20133 Milano, Italy

[3]Dipartimento di Scienza dei Materiali, Università di Milano-Bicocca, Via Cozzi 53, I-20125 Milano, Italy

[4]European Theoretical Spectroscopy Facility (ETSF), Via Celoria 16, 20133 Milano, Italy

[5]Actinium Chemical Research, Via Casilina 1626/A, I-00133 Rome, Italy

[6]Istituto Nazionale di Astrofisica. Osservatorio Astrofisica di Catania,

Via S. Sofia 78, I-95123 Catania, Italy

[*]    Corresponding authors. E-mail: luca.ravagnan@mi.infn.it (L.R.)
[+]    Present address: Centre for Atomic-scale Materials Design (CAMD), Department of Physics, Technical University of Denmark, DK-2800 Kongens Lyngby, Denmark

## ABSTRACT:

We perform a systematic investigation of the resonance and vibrational properties of naphthyl-terminated *sp* carbon chains (dinaphthylpolyynes) by combined multi-wavelength resonant Raman spectroscopy (MWRR), ultraviolet-visible spectroscopy, and Fourier-transform infrared spectroscopy (FT-IR), plus *ab-initio* density functional theory (DFT) calculations. We show that MWWR and FT-IR spectroscopy are particularly suited to identify chains of different lengths and different terminations, respectively. By DFT calculations we further extend those findings to *sp* carbon chains end-capped by other organic structures. The present analysis shows that combined MWRR and FT-IR provide a powerful tool to draw a complete picture of chemically stabilized *sp* carbon chains.

**KEYWORDS**: *sp*-carbon chains, Raman spectroscopy; Infrared spectroscopy; ab-initio calculations.

1.  Introduction

The interest about *sp* carbon chains (*sp*CCs) as building blocks for carbon-based nanoelectronics is very recent [1,2], *sp*CCs have been considered for a long time as objects relevant only for astrochemistry [3-6]: this is due to their high reactivity and instability supporting the opinion that only conditions as those typical of circumstellar and interstellar clouds could allow the presence of large quantities of linear carbon chains acting as intermediates for the formation of either fullerenes or poly-aromatic hydrocarbons (PAHs) [7,8].

In the last decade, an increasing amount of experimental evidences have demonstrated that *sp*CCs can be mass produced in standard laboratory conditions, and, more surprisingly that *sp*CCs can coexist with other carbon allotropic forms [9-10], with *sp*CCs being stabilized by different end-capping groups and in particular by graphene terminations [11-13]. Graphene-terminated *sp*CCs exhibit remarkable magnetic properties, sensitivity to axial strain and rectifying performances potentially enabling their use as spin-filters and spin-valves in magnetic nanodevices [14-17]. End-capped *sp*CCs stabilized by different terminations have also been proposed as building blocks of hydrogen-storage materials [18-19].

Despite the increasing amount of theoretical work dealing with end-capped *sp*CCs, relatively few experimental studies elucidating the structural and functional properties (chain length, bonding nature, electronic structure) are available. Vibrational and optical spectroscopies can provide information about the physical and structural properties of end-capped *sp*CCs, but the identification of the spectroscopic signatures of specific structures is, until now, an open issue.

Recently we demonstrated the possibility to produce efficiently *sp*CCs of different lengths terminated by naphtyl groups (dinaphthylpolyynes) [20]. These compounds represent a family of chemically stabilized α,ω-diarylpolyynes of high interest for several reasons: on one side, they are potentially useful in optoelectronic applications being promising building blocks for the production of $sp$-$sp^2$ carbon systems for nano-electronic devices; on the other side, they constitute the simplest examples of *sp*CCs terminated by PAHs that could be identified in the interstellar medium (naphthalene is indeed one of the simplest and most abundant PAHs).

Here we present a full vibrational characterization, both experimental and theoretical, of dinaphthylpolyynes [20]: our approach combines multi-wavelength resonant Raman (MWRR) and Fourier-transform infrared (FT-IR) spectroscopic characterization, with density-functional theory (DFT) modeling. Thanks to the combination of these approaches we obtain an overall description of the vibrational properties of dinaphthylpolyynes thus providing a complete framework for the interpretation of vibrational data of end-capped *sp*CCs.

**2. Methods**

*2.1 Synthesis and chromatographic characterization of dinaphthylpolyynes*

The synthesis of a series of dinaphthylpolyynes (in short represented as Ar-C$_{2n}$-Ar, with Ar = naphthyl group and $n$ = number of acetylenic units of the chain) was obtained, as described in detail in ref. [20], by reacting copper(I)-ethynylnaphthalide with diiodoacetylene under the Cadiot-Chodkiewicz reaction conditions [21,22]. The obtained molecules are dissolved in decahydronaphthalene, and were characterized by high-performance liquid chromatograph (HPLC) analysis using an Agilent Technologies 1100 station with C$_8$ column (using acetonitrile/water 80/20 as mobile phase at a flow rate of 1.0 ml/min). The HPLC station was also equipped with a diode array detector, allowing recording the ultraviolet-visible (UV-vis) absorption spectra of the separated molecules present in the solution [20]. The UV-vis absorption spectrum of the whole solution was also acquired by using a UV-vis Jasco 7850 spectrometer.

*2.2 Multi-wavelength Raman and Infrared characterization*

MWRR spectroscopy characterization of the dinaphthylpolyynes solution was performed by using seven different laser wavelengths produced by an Ar ion and a He-Cd laser (Spectra Physics, BeamLok series 2065-7 and Kimmon, IK5352 R-D series, respectively). The wavelengths used were 514.5 nm, 496.5 nm, 488.0 nm, 476.5 nm, 457.9 nm and 363.8 nm for the Ar ion laser, and 441.6 nm for the He-Cd laser. For each wavelength, bandpass clean-up filters (MaxLine filters from Semrock) were used in order to remove spurious plasma lines produced by the lasers.

MWRR spectra were acquired by using a single-grating spectrometer (Acton SP-2558-9N with a 1200 groves/mm grating, blazing 600 nm) equipped with a liquid nitrogen cooled CCD camera (Roper-Princeton Instruments Spec10:400B/LN) and notch filters in order to remove the Rayleigh scattering (StopLine from Semrock). Thanks to this setup the spectra were recorded in a unique acquisition window (typically comprised between 1000 $cm^{-1}$ and 2500 $cm^{-1}$), repeating 100 times a 60 seconds acquisition. By this approach it was both possible to obtain spectra with high signal to noise ratio (by averaging the different acquisitions) and to detect any possible degradation of the solution induced in time by the laser irradiation. Furthermore, this setup allows us to switch from one laser wavelength to the next in a very short time interval, simply by changing the bandpass and notch filters, and moving the monochromator grating.

Fine calibration of the Raman shift of the acquired spectra was achieved by using the approach proposed by N.C. Craig and I.W. Levin [23]. After each measurement, a spectrum of the plasma lines produced by the lasers was collected by removing the bandpass clean-up filter in the optical system. The accurately tabulated plasma line positions allowed us to correct any possible offset or non linearity of the monochromator and to control the uncertainty on the wave-number axis within 1 $cm^{-1}$ for all excitation wavelengths used (also checked by measuring the G-band position at 1581 $cm^{-1}$ in the spectrum of a highly oriented pyrolytic graphite sample).

The infrared spectrum of the dinaphthylpolyynes solution was acquired in transmission mode by a Perkin Elmer FT-IR/F-FIR (far infrared) Spectrometer "Spectrum 400" with a Universal Attenuated Total Reflectance (UATR) attachment and liquid $N_2$ cooled Mercury Cadmium Telluride detector operating in the range 750-2500 $cm^{-1}$ with a 2 $cm^{-1}$ resolution. The spectrum of the pure decahydronaphtalene was also acquired and then subtracted from the spectrum of the whole solution in order to remove its infrared features.

*2.3 DFT simulations*

We simulate isolated (gas phase) dinaphthylpolyynes with *n*=2-8 acetylenic units within the framework of DFT [24,25] in the local spin-density approximation (LSDA).

We use the QUANTUM ESPRESSO package [26], based on a plane-wave basis (40 Ryd kinetic energy cutoff) and ultra-soft pseudo-potentials [27,28].

We relax the atomic position until the largest residual force and energy difference are smaller than 8 pN and $10^{-5}$ Ryd, respectively. Starting from the relaxed position, we then compute the molecular vibrational modes, plus their infrared intensities and non resonant-Raman cross-sections, using density-functional perturbation theory [26]. The computational details are similar to those validated and used, e.g., in Refs. [1, 20] for the simulations of the structure and vibrational properties of *sp*CCs terminated by $C_{20}$, graphene nanoribbons (NR) and $CH_2$.

# 3 Results and Discussion

*3.1 Raman-active modes*

Figure 1a reports the electronic absorption spectrum of each molecular species contained in the dinaphtylpolyynes mixture, measured by using the diode array detector interfaced to the column of the HPLC. Table 1 reports the relative molar concentration *C(n)* of the individual components, as estimated by means of the Lambert-Beer law [20].

| $n$ | dinaphthylpolyyne Formulas | $C(n)$ |
|---|---|---|
| 2 | Ar-$C_4$-Ar | 47 ± 4 % |
| 3 | Ar-$C_6$-Ar | 19.7 ± 0.6 % |
| 4 | Ar-$C_8$-Ar | 18.2 ± 2.2 % |
| 5 | Ar-$C_{10}$-Ar | 11.3 ± 1.2 % |
| 6 | Ar-$C_{12}$-Ar | 4.1 ± 0.3 % |
| $n>6$ | | Not detected |

Table 1. Relative molar concentration of the dinaphthylpolyynes (with 2≤*n*≤6) in the solution [20].

One can clearly observe typical multi-peak structure in the optical absorption spectra due to Franck-Condon satellites of the main electronic transition [20, 29-32]. The wavelength of the purely electronic transition, i.e. the longest-wavelength transition (LWT), shifts linearly with the number of acetylenic units, due to the hybridization of the molecular orbitals of the *sp*CCs with those of the terminations [20].

Figure 1a also reports (as vertical lines) the positions of the individual laser wavelengths used for MWRR spectroscopy. As can be observed, the two laser lines at longer wavelength (514.5 and 496.5 nm) do not match any absorption line of the spectra shown, while from 488.0 nm downwards, at least one absorption peak is crossed. At 363.8 nm, all the dinaphthylpolyynes peaks are crossed.

The matching of the laser wavelength with the energy of an electronic absorption of a molecular species has important consequences for the obtained Raman spectra [33]. Indeed, although Raman scattering occurs for any excitation wavelength, when the energy of the incident photon matches the energy of an optical absorption peak (i.e. an electronic transition) of a species present in the solution, a strong enhancement of the Raman intensity corresponding to this particular species occurs [34]. In contrast, resonance does not affect the position of the Raman peaks (i.e. the vibrational frequency) but only their intensity [32, 34].

Figure 1b shows the Raman spectra obtained on the dinaphthylpolyynes solution using the 7 excitation wavelengths: all the spectra are characterized by a multi-peak structure. In particular, seven of those peaks (indicated as P1-P7 in Figure 1b, and summarized in Table 2) are present in all spectra at the same wavenumber, but with different intensities.

| Peak label | Raman shift | Dinaphthylpolyynes assignment |
|---|---|---|
| P1 | 1957 ± 4 | |
| P2 | 1980 ± 3 | |
| P3 | 2005 ± 2 | Ar-$C_{16}$-Ar (?) |
| P4 | 2029 ± 6 | Ar-$C_{14}$-Ar (?) |
| P5 | 2057 ± 6 | Ar-$C_{12}$-Ar |
| P6 | 2090 ± 3 | Ar-$C_{10}$-Ar |
| P7 | 2129 ± 2 | Ar-$C_8$-Ar |

Table 2. Assignment of the experimental Raman peaks to dinaphthylpolyynes of different lengths.

Considering for instance the peak P5 in Figure 1b, its intensity remains substantially unchanged when acquired with the 514.5 nm and 496.5 nm laser wavelengths, while it becomes much stronger at 488.0 nm, and reaches its maximum intensity at 476.5 nm. It then drops in intensity at 457.9 nm and increases again at 441.6 and 363.8 nm. The electronic absorption spectra of Figure 1a clarifies that this behavior follows exactly the evolution of the optical absorption of Ar-$C_{12}$-Ar. It is thus possible to attribute, based on its resonant behavior, the peak P5 to Ar-$C_{12}$-Ar. Following the same approach, also the peaks P6 and P7 have resonant behaviors that can be clearly attributed to Ar-$C_{10}$-Ar and Ar-$C_8$-Ar respectively. Also the peaks P1, P2, P3 and P4 show resonant behaviors, which however do not correspond to any of the dinaphthylpolyynes identified by HPLC. In particular their intensity indicates that they are due to species resonating already at 514.5 nm, and thus at a wavelength exceeding that of the LWT of any of the identified species.

In order to obtain a better understanding, absorption spectroscopy measurements were repeated on the whole solution, before being separated in the HPLC column. Figure 2 shows the resulting spectrum, compared to the sum of the individual spectra, reported in Figure 1a, of the single-$n$ Ar-$C_{2n}$-Ar molecules (weighted with the relative molar concentration of Table 1). The sum of the spectra, although significantly broader, reproduces accurately the spectrum of the whole solution. However as highlighted in the inset of Figure 2, an extra weak peak (between 490 and 530 nm) is present in the whole-solution spectrum, that is absent in the sum spectrum. This peak must be related to the dinaphthylpolyynes with $n>6$, as for instance Ar-$C_{14}$-Ar and Ar-$C_{16}$-Ar, that are contained in extremely low quantity in the solution and that were thus not identified by HPLC [20]. Indeed, considering the linear evolution of the LWT observed for the dinaphthylpolyynes with $n$=2-6, one would expect for $n$=7 a LWT centered at about 500 nm, while for $n$=8 a value for LWT of about 525 nm [20], consistently with the position of the observed peak. Remarkably, the resonant evolution of P4 for the 514.5 nm, 496.5 nm and 488.0 nm laser lines in Figure 1b follows nicely the evolution of this weak absorption peak. Indications of the presence of dinaphthylpolyynes with $n>6$ in the solution are furthermore supported by the DFT simulations. Indeed, as typical for *sp* linear carbon chains, the Raman active mode of the dinaphthylpolyynes bearing the strongest Raman intensity shows a displacement pattern localized near the chain center (for more details about the pattern of longitudinal atomic displacements, see figure reported in the supplementary materials [35]). This so-called Raman-α mode (R-α) [1, 36] occurs in the range 1950 ÷ 2300 cm$^{-1}$ and is characterized by its strong dependence on the *sp*CC length, red shifting almost uniformly as this length increases (see Figure 3a). The attribution based on the resonant behavior of P5, P6 and P7 to the alpha mode of dinaphthylpolyynes Ar-$C_{2n}$-Ar with $n$ = 6, 5 and 4 respectively is also supported by the simulated frequencies, which match the experimental peaks within 15 cm$^{-1}$ (i.e. 1%, see Table 2). Furthermore, the simulations support the attribution of the peaks P1 - P4 at smallest Raman shifts to $n>6$ dinaphthylpolyynes.

None of the P1-P8 peaks in the Raman spectra can be associated to $n$ = 2 and 3 dinaphthylpolyynes (i.e. Ar-$C_4$-Ar and Ar-$C_6$-Ar), even though, altogether, they represent almost 70% of the *sp*CCs present in solution (see Table 1). Only in the Raman spectrum acquired with the 363.8 nm laser line four peaks (P8-P11 in Figure 1b, not present in the other spectra) are clearly identifiable: these peaks have positions compatible with the vibrational *sp*CCs modes of Ar-$C_4$-Ar and Ar-$C_6$-Ar. The absence of these peaks in the other spectra can be understood by considering the computed non-resonant Raman cross section for the R-α modes shown in Figure 4a. Indeed, the Raman cross section grows rapidly with the increasing chain length, being for example 20 times larger for $n = 6$ than for $n = 2$. Figure 5 shows the comparison between the experimental Raman spectra recorded using the 514.5 nm laser line (which is non-resonant for the dinaphthylpolyynes with $n = 2$-6) and the spectrum obtained by summing the simulated non-resonant Raman spectra of the dinaphthylpolyynes weighted by their relative molar concentrations (Table 1). The experimental and simulated spectra display a remarkable agreement. In particular, the simulated spectrum confirms that the Raman peaks from chains with $n = 4$-6 dominate over those relative to $n = 2$ and $n = 3$. The simulations also predict the small spectral features P5' and P6', thus attributed to secondary Raman active modes (weaker than the main α modes).

In the simulated spectrum of Figure 5b we include the contributions of $n = 7$ and 8 *sp*CCs. Those peaks (indicated as dashed peaks in Figure 5b) were introduced by assuming arbitrary relative molar concentrations of 1% and 0.5% respectively (since $n = 7$ and 8 were not detected by HPLC). On one hand, their positions are consistent with the positions of P4 and P3, on the other hand their observed intensity is most likely enhanced by resonance: indeed, as previously discussed, Ar-$C_{14}$-Ar and Ar-$C_{16}$-Ar are expected to have electronic transitions matching the 514.5 nm laser line (see also Figure 2). Following the same reasoning, P1 and P2 might be related to even longer dinaphthylpolyynes (as for example Ar-$C_{18}$-Ar and Ar-$C_{20}$-Ar) that although present in negligible concentrations in the solution (far below the detection limit of the HPLC), appear in the Raman spectrum due to their extremely large Raman cross sections, further enhanced by resonance.

Likewise, we identify the P8-P11 peaks with the simulated vibrational modes of Ar-$C_4$-Ar and Ar-$C_6$-Ar because they can be observed only when using the 363.8 nm excitation line (Figure 1b) to match the optical resonance (Figure 1a) and therefore to compensate for the small cross-sections of these short chains.

In summary, by combining electronic absorption spectra recorded during HPLC, MWRR and DFT-LSDA simulations it is possible to fully and coherently describe the mixture of *sp*CCs of several lengths. In particular we can identify the vibrational modes of the different species present in the system, whose position is strongly dependent on the chain length. Furthermore we demonstrate that Raman spectroscopy with long excitation wavelengths is highly sensitive to long *sp*CCs (rather than to short chains), due to their larger non-resonant cross sections, further enhanced by selective resonant absorption associated to their smaller HOMO-LUMO gaps. As the laser wavelength is shifted toward the UV, shorter *sp*CCs becomes progressively resonant, with their R-α peaks becoming the dominant features of the spectra. This resonance mechanism gives a chance to the R-α peaks of the abundant $n = 2$ and $n = 3$ *sp*CCs to become visible despite their extremely small Raman cross sections.

### *3.2 Infrared-active modes*

Figure 6a shows the FT-IR spectra measured on the dinaphthylpolyynes solution. Similarly to MWRR, a multi-peak structure can be clearly identified, which is distributed on a much narrower wavenumber interval (about 100 cm$^{-1}$ compared to the 250 cm$^{-1}$ interval of Raman).

The peak positions agree well with the IR frequencies of dinaphthylpolyynes with $n = 1\text{-}6$ as measured by K. Nakasuji et al. in ref. [31]. Nevertheless, the small separation between the different peaks in Figure 6a (below 20 cm$^{-1}$) makes their precise assignment to the dinaphthylpolyynes of specific chain length quite problematic. The DFT-LSDA simulations help us understand the small dispersion of the IR modes. The displacement patterns of the most intense IR-active modes of the dinaphthylpolyynes involve two main patterns: one (named IR-α) characterized by a displacements distinctly localized at the chain center is present ubiquitously for all dinaphthylpolyynes; another one (named IR-β) with a bi-lobed pattern occurs from $n > 5$. As shown in Figure 3b, the IR-α mode vibrates at a frequency almost independent of the length of the dinaphthylpolyynes (with the exception of the $n = 2$ chain), while the IR-β red-shifts uniformly as the chain length increases, like in the case of the R-α mode. Although the IR-α mode is always present, it carries the main intensity only up to $n = 5$, while it becomes very weak starting from n = 6, where the IR-β mode "steals" most of the dipole intensity (see Fig. 4b). Nevertheless, for $n > 6$ the intensity of the IR-α mode remains relevant, bearing approximately 20% of the intensity of the C=C modes for $n = 8$. As illustrated in Figure 4b (IR-α for $n=6$ is less than $10^{-4}$), the cross section of the most intense IR mode (the IR-α up to $n = 5$, and the IR-β for $n > 5$) increases exponentially like the R-α mode, showing that also IR spectroscopy is much more sensitive to long chains than to short ones. Unfortunately, while Raman spectroscopy, taking advantage of optical resonance, can be tuned to enhance the response of one molecular species, this cannot be done with IR spectroscopy.

Figure 6b shows the simulated IR spectrum obtained, as in the case of Raman in Figure 5b, by summing the simulated IR spectra of the dinaphthylpolyynes weighted by their relative molar concentrations. It must be noted that, in order to match the experimental spectrum in Figure 6a, a global 50 cm$^{-1}$ red-shift of the simulated isolated-molecule spectrum was required. Although shifts of this entity between the simulated frequencies and the experimental one are common (see for example ref. [36,37]), this circumstance becomes surprising when considering the very good agreement of the simulated Raman modes with experiment, with no need of any shift (Figure 5). Nevertheless, one can understand this by considering that, contrary to the R-α mode, the most intense IR modes have wide bond displacement near the chain termination (i.e. at the naphthyl groups). As a consequence, it is likely that the interaction between the dinaphthylpolyynes and the solvent (attenuating the motion of the chain termination) induces the observed softening of the IR active modes much more than the R-α mode, which are localized near the chain center. Similar effects have been observed by many authors also for other systems both in solution and in the solid state [37-41]. Although the solvent-induced shift need not be uniform for dinaphthylpolyyne of different lengths, and also considering the frequent tendency of LDA to overestimate the vibrational frequencies [42-44], the overall agreement between the shifted simulated spectrum in Figure 6b and the experimental one in Figure 6a is good.

In particular, we ascribe the accumulation of the components in the 2150-2200 cm$^{-1}$ interval to the lack of dispersion of the IR-α mode and the weak dispersion of the IR-β as a function of the chain length.

*3.3 From dinaphthylpolyynes to general properties of end-capped spCCs*

In order to verify the general validity of the observed behavior for the Raman and IR modes on dinaphthylpolyynes, we also carried out DFT-LSDA calculations of *sp*CCs end-capped by other carbon based structures: $C_{20}$, graphene nanoribbons (NR), and $CH_2$ [1]. Figure 7 shows the resulting simulated R-α and IR-α frequencies as functions of the number of acetylenic units. Figure 7 also reports the bond-length alternation of the chains (BLA, see definition in ref. [1]) and the length of the first bond between the chain and the terminating structure (FBL).

Figure 7 clarifies that the BLA and FBL values allow to characterize precisely the nature of the different *sp*CCs. Specifically, the BLA measures of the overall degree of dimerization of the *sp*CC. As discussed in Ref. [1], although the BLA is maximum for "ideal" polyynes (having alternating single-triple bonds) and negligibly small for cumulenes (having all double bonds), its value is not only affected by the nature of the chain termination (being inversely related to the order of the bond between the chain and the terminating group), but also by the chain length and, for $sp^2$-terminated chains, by the chain torsion [1]. Indeed, as shown in Figure 7, for each termination type the BLA decreases progressively with increasing *n*. In contrast, the FBL provides a direct indication of the bond strength between the terminating group and the chain, and its value is almost independent of the chain length.

Consider now the *n*-dispersion of the R-α and IR-α modes: the general validity of the trends of dinaphthylpolyynes discussed in the Sect. 3.1 and 3.2 appears clear. Indeed, the linear *n*-dispersion of the R-α mode and the approximate *n*-independence of the IR-α mode (for *n* > 2) are observed for all kinds of termination. Remarkably, as shown in Figure 7, the evolution of the R-α and IR-α modes follows, respectively, the dispersion of the BLA and FBL values.

Note that the R-α frequency depends only weakly on the nature of the terminating group, but, similar to the BLA, it is dominated by the strong dispersion with *n*. Conversely, the frequency of the IR-α mode (for *n* > 2) correlates to the value of the FBL and, in turn, to the nature of the chain termination. Those observations strongly indicate that Raman and IR characterization can provide complementary information*: Raman identifies the presence of spCCs of different length within a given family* (i.e. with a specific terminating group), while *IR distinguishes between different families of spCCs compounds regardless their length distribution.*

## 4. Conclusions

We performed an experimental and theoretical characterization of the vibrational properties of Naphthyl-terminated *sp*CCs, by combining MWRR and FT-IR spectroscopies with theoretical DFT-LSDA modeling. One R-α mode dominates the Raman spectra of dinaphthylpolyynes, while the IR spectrum contains one dominant IR-α dipole-active mode up to *n*=5, plus a second IR-β from *n*>5 onwards; both R-α and IR-β red shift as the chain length increases, while IR-α is almost unaffected by the chain length. As a consequence the non-resonant Raman spectrum of a mixture of terminated *sp*CCs of different lengths covers almost uniformly the whole frequency range of *sp* carbon (1900-2300 $cm^{-1}$), while, due to the length-independence of IR-α, the IR spectrum exhibits an accumulation of components around a single frequency, here 2200 $cm^{-1}$ for dinaphthylpolyynes, with only few weak features far from this frequency.

We demonstrated that resonance effects can be exploited with MWRR spectroscopy for the identification of the detected peaks as R-α modes of dinaphthylpolyynes of specific lengths. Raman spectroscopy is far more sensitive to long chains rather than short ones, due to: (i) the rapid increase of the non-resonant Raman cross section with chain length and (ii) the onset of resonant enhancement already with the long-wavelength (visible) excitation line following the decreasing HOMO-LUMO gap with increasing chain length [20,45].

Due to the weak *n*-dispersion, no complete assignment of the individual peaks to specific IR-α and IR-β modes was feasible. Nevertheless, DFT calculations indicate that the IR modes, much more than Raman ones, are strongly affected by the chain termination groups, rigidly shifting to lower wavenumbers as the strength of the terminal bonds increases.

The combined characterization of end-capped *sp*CCs by MWRR and IR spectroscopy (complemented by DFT calculations) allowed us to propose a paradigm for the full identification of general *sp*CCs-containing compounds: MWRR spectroscopy allows the identification of different chain lengths, while IR allows distinguishing families of chains with different terminations (if present). The combination of the two techniques represents therefore a very powerful tool for the characterization of complex carbon-based materials [46-49].

**Acknowledgments**

The authors wish to thank W. R. Browne from the "Stratingh Institute for Chemistry," University of Groningen and O. Ivashenko from the "Zernike Institute for Advanced Materials," University of Groningen for the precious help in the FT-IR measurements.

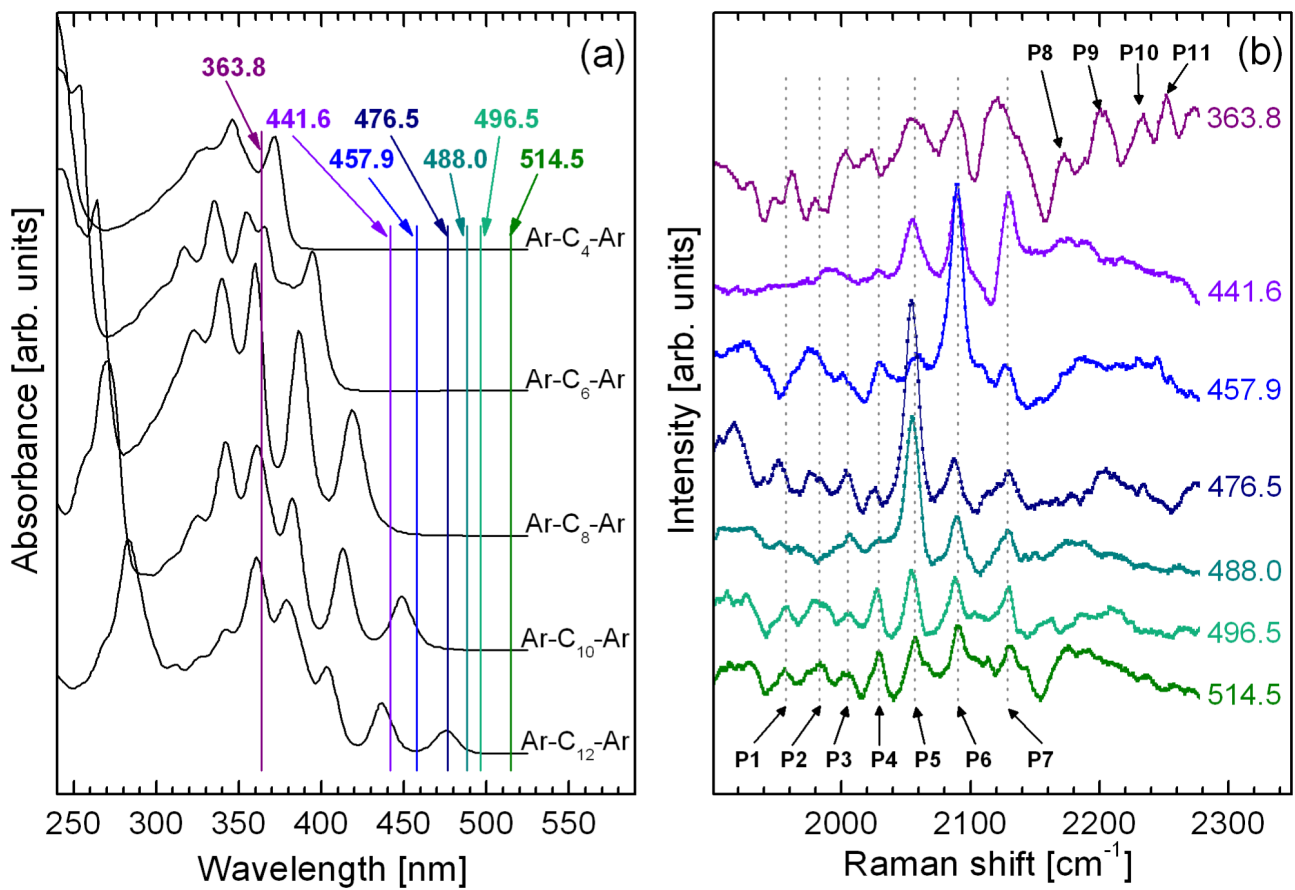

Figure 1 (a) UV-vis absorption spectra of dinaphthylpolyynes of different chain length measured by using the diode array detector interfaced to the column of the HPLC. Vertical lines represent the excitation wavelengths (i.e the laser lines) used for MWRR spectroscopy. (b) MWRR spectra of the whole solution obtained using 7 different excitation wavelengths, whose value in nm is indicated at the right.

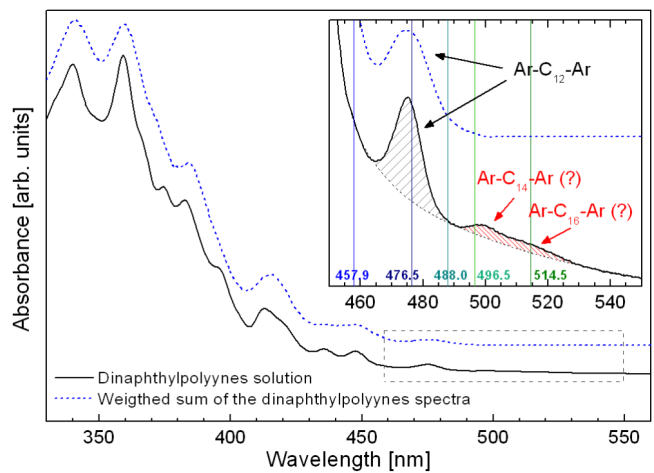

Figure 2 Comparison between the UV-vis spectrum of the solution containing the mixture of dinaphthylpolyynes with different chain length (black solid line) and the weighted sum of the spectra recorded on the individual dinaphthylpolyynes separated by HPLC from the same solution (blue dashed line). Inset: a blowup of the 450-550 nm region, with several excitation lines used for the Raman experiments, and the peak assignments.

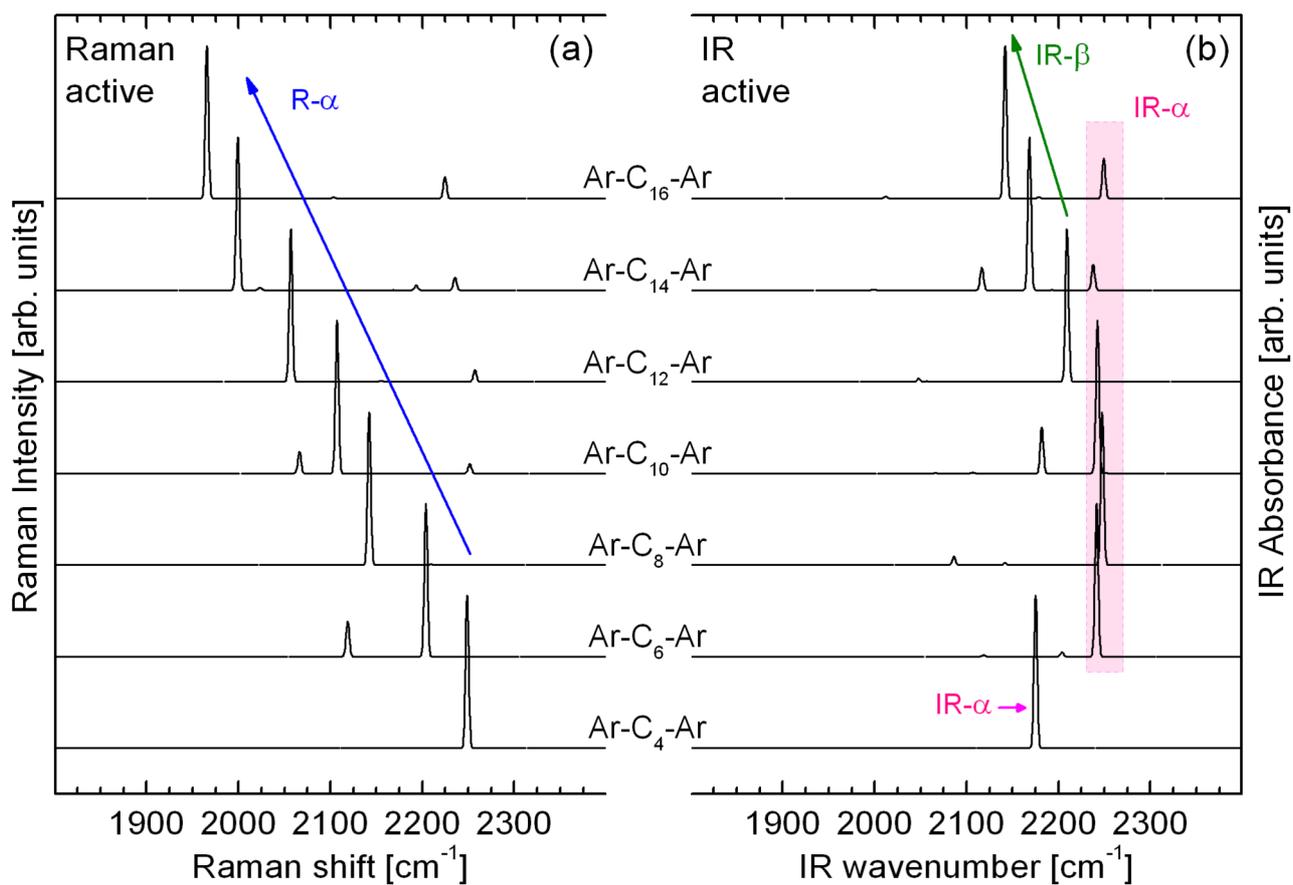

Figure 3: The DFT-simulated frequencies of the (a) Raman and (b) IR active modes of dinaphthylpolyynes. The intensities are normalized to the most intense mode in the 1900-2300 cm$^{-1}$ window.

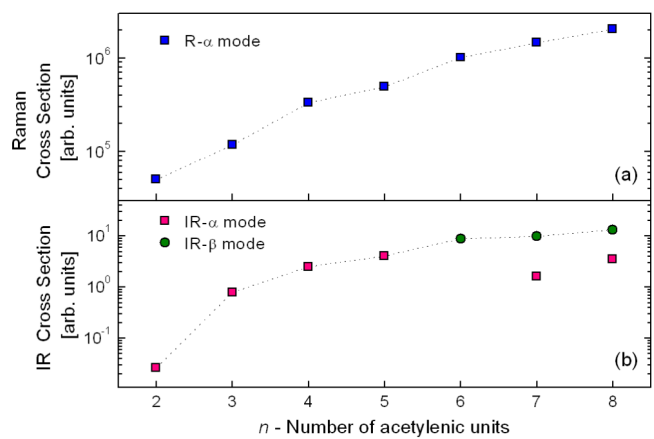

Figure 4: (a) Non-resonant Raman cross sections of the R-α mode of dinaphthylpolyynes as a function of the chain length; (b) the dipole intensity of the IR-α and IR-β modes of dinaphthylpolyynes (IR-α for $n = 6$ is less than $10^{-4}$).

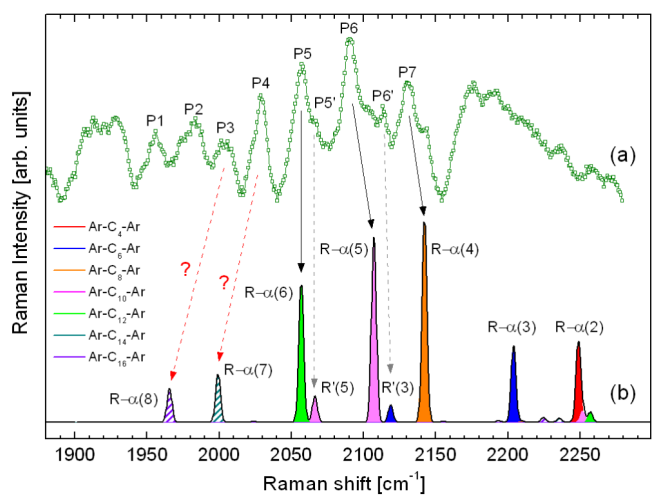

Figure 5: (a) Experimental Raman spectrum measured using the 514.5 nm excitation wavelength. (b) Simulated Raman spectrum obtained by combining the DFT spectra of Fig. 3(a) with the computed non-resonant cross-sections of Fig. 4(a), weighted by the measured concentrations. Arrows indicate the correspondence between experimental and computed peaks.

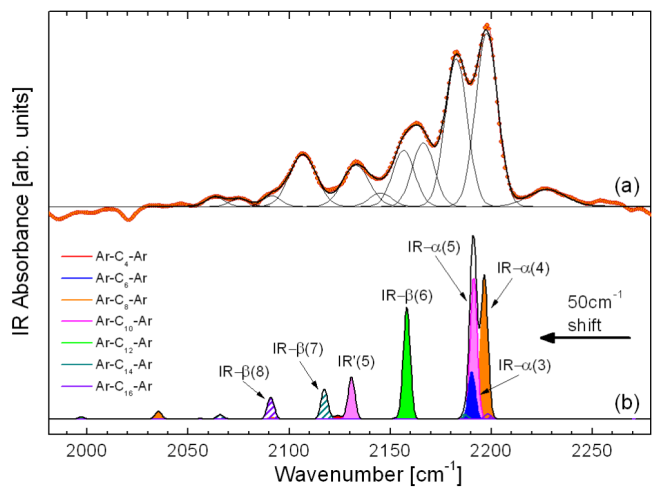

Figure 6: (a) Measured infrared spectrum of the solution containing Ar-$C_{2n}$-Ar molecules (small circles), with its decomposition into Gaussians (thin solid). (b) Computed infrared spectrum obtained by combining the DFT spectra of Fig. 3(b) (red-shifted by 50 cm$^{-1}$) and dipole intensities of Fig. 4(b), weighted by the measured concentrations.

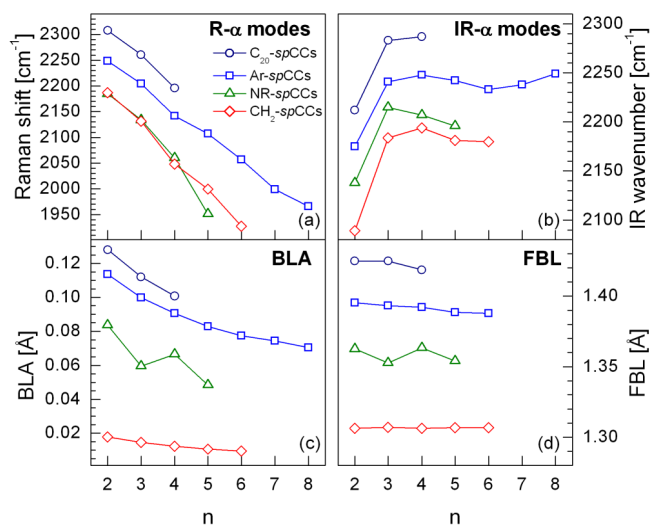

Figure 7: Simulated (a) R-α frequencies, (b) IR-α frequencies, (c) BLA [1], and (d) the length of the first bond between the chain and the terminating structure, as functions of the number of acetylenic units forming the *sp*CC bonded to: $C_{20}$ clusters (circles), naphtil groups (squares), graphene nanoribbons (triangles), and $CH_2$ groups (diamonds).